\documentclass[twocolumn,aps,showpacs,prb,amsmath,amssymb,floatfix,superscriptaddress]{revtex4-2}

\usepackage{graphicx}
\usepackage{microtype}
\usepackage{siunitx}
\usepackage[version=4]{mhchem}
\usepackage{xcolor}
\usepackage[normalem]{ulem}

\usepackage[color=orange!60,textsize=footnotesize]{todonotes}
\DeclareSIUnit\angstrom{\text{\AA}}
\DeclareSIUnit\bar{bar}

\usepackage{pdfpages}
\usepackage{pgffor}
\usepackage[hidelinks]{hyperref}
\makeatletter
\AtBeginDocument{\let\LS@rot\@undefined}
\makeatother
\def\supplementfilename{supplement}
\newif\ifarXiv
\arXivtrue

\begin{document}

\title{Giant gas permeance through a nanoporous crystalline 2D polymer}

\author{Jonathan D'Emidio}
\thanks{These authors contributed equally to this work.}
\affiliation{Department of Physics and Astronomy, University of Tennessee, Knoxville, 1413 Circle Drive, Knoxville, Tennessee 37996, USA}

\author{Daniel J. Murray}
\thanks{These authors contributed equally to this work.}
\affiliation{Department of Chemistry, University of Nevada, Reno, 1664 N. Virginia St, Reno, Nevada 89557, USA}
\affiliation{Nuclear Science and Technology, Idaho National Laboratory, 1955 Fremont Ave, Idaho Falls, Idaho 83415, USA}

\author{William B. Thompson}
\thanks{These authors contributed equally to this work.}
\affiliation{Department of Chemistry, University of Nevada, Reno, 1664 N. Virginia St, Reno, Nevada 89557, USA}

\author{Jarvis C. Humlick}
\affiliation{Department of Chemistry, University of Nevada, Reno, 1664 N. Virginia St, Reno, Nevada 89557, USA}

\author{K. C. Schwab}
\affiliation{Thomas J. Watson Sr. Laboratory of Applied Physics, California Institute of Technology, 1200 E. California Blvd., Pasadena, California 91125, USA}

\author{Adrian Del Maestro}
\email[Corresponding author: ]{Adrian.DelMaestro@utk.edu}
\affiliation{Department of Physics and Astronomy, University of Tennessee, Knoxville, 1413 Circle Drive, Knoxville, Tennessee 37996, USA}
\affiliation{Min H. Kao Department of Electrical Engineering and Computer Science, University of Tennessee, Knoxville, 1520 Middle Drive, Knoxville, Tennessee 37996, USA}

\author{Benjamin T. King}
\email[Corresponding author: ]{king@unr.edu}
\affiliation{Department of Chemistry, University of Nevada, Reno, 1664 N. Virginia St, Reno, Nevada 89557, USA}

\begin{abstract}
Chemical separation processes represent one of the largest contributors to industrial energy consumption, underscoring the need for the most efficient separation strategies \cite{Sanders:2013}. Many separation processes rely on selective transport through porous media \cite{Yuan2022}. In this context, two-dimensional (2D) materials offer a promising route to atomically precise porosity beyond the reach of traditional top-down fabrication \cite{Colson:2013natchem}, with the potential to enhance separation performance by increasing flow \cite{Wang:2017fu}. Here we report measurements of gas transport through a one-molecule-thick layer of suspended, nanoporous poly(carboxyfantrip), a 2D polymer that has identical, nanometer-sized pores that cover nearly a quarter of the surface \cite{Kissel:2014nc,muller2017structural}. This combination of ultimate thinness with high areal pore density gives normalized gas flows (permeances) that approach the effusion limit, three to four orders of magnitude above gas-sieving graphene membranes. We find that gases spanning nearly two orders of magnitude in molecular weight (H$_2$, He, N$_2$, CO$_2$, and SF$_6$) obey Graham’s law and approach the ideal limit of non-interacting ballistic transport, despite strong, species-dependent gas–pore interactions. Simulations reproduce the measured permeances within experimental uncertainty and provide an atomistic picture of the transport process. This work demonstrates the potential of synthetic crystalline 2D polymers for chemical separation, water filtration, and quantum sensing.
\end{abstract}

\maketitle

From respiration to dialysis, from natural gas purification to desalination, membrane transport and separation processes are technologically ubiquitous \cite{Sanders:2013}.  In membrane separations, mass transport rates depend on membrane thickness, with thinner membranes exhibiting faster transport \cite{Wang:2017fu}.  However, the thinnest possible membranes such as graphene are essentially impervious to even the lightest gases with energy barriers predicted to be as high as $\sim$\SI{20}{\electronvolt} \cite{Hrusak:1992ai,Bunch:2008cv,Leenaerts:2008ff}. As a result, graphene-based membranes for separation must rely on top-down approaches in which pores are introduced through the formation of defects.

Top-down pore formation can be achieved using techniques such as focused ion beams \cite{Celebi:2014sc}, UV-induced oxidative etching \cite{Koenig:2012nn}, or controlled dielectric breakdown \cite{Kwok2014,Kuan:2015apl}. Computational \cite{Jiang:2009nl,Yuan:2017acsnano} and experimental \cite{Koenig:2012nn,Celebi:2014sc} studies of gas transport through such porous graphenes have demonstrated that atomically thin membranes can achieve permeances far exceeding those of conventional polymer membranes \cite{Yuan2022}.    However, the top-down approach faces two significant practical challenges.  First, achieving a useful total flux requires a large number of pores, and creating such pore density remains challenging.  Second, top-down etching methods produce a statistical distribution of pore sizes and shapes that reduces selectivity in gas separations~\cite{Villalobos2022}.

An alternative strategy is to use intrinsically porous 2D materials, in which the pores are incorporated into the structure during synthesis, rather than created afterward.  Such materials include 2D covalent organic frameworks, metal-organic frameworks, and synthetic 2D polymer crystals, which together comprise a rich design space for porous 2D sheets. This bottom-up strategy has enabled the synthesis of several intrinsically porous 2D materials \cite{Michl:2002pnas,Schlueter:2020natrev,Feng:2012anie,Colson:2013natchem,Bhola:2013jacs,Kissel:2014nc}. In contrast to top-down methods, the pores in a 2D polymer crystal are identical at the atomic scale and present at an extremely high area density. 

Poly(carboxyfantrip), the subject of this study, is a synthetic 2D polymer crystal that demonstrates these advantages.  This material is prepared by [4 + 4] photocycloaddition of the trigonal, star-shaped carboxyfantrip monomer that has been pre-organized into a crystalline honeycomb monolayer at the air--water interface \cite{muller2017structural,muller2018two,Murray:2016thesis}, following earlier work on a similar amphiphilic monomer \cite{Murray:2015fe}. The resulting polymer is $\sim$\SI{1.3}{\nano\meter} thick and adopts a distorted hexagonal lattice ($a = \SI{1.71}{\nano\meter}$, $b = \SI{1.92}{\nano\meter}$, $\gamma = \SI{70}{\degree}$) in which every pore is structurally identical, with a pore density of $\SI{3.2E{17}}{\meter^{-2}}$ \cite{Kissel:2014nc}. A rough estimate based on atomic kinetic radii indicates that 30\% of the membrane area is open pore. Free-standing membranes of poly(carboxyfantrip) are mechanically robust: they have been transferred from the air--water interface onto substrates with arrays of holes as large as $ \SI{100}{\micro\meter} \times \SI{100}{\micro\meter}$ \cite{muller2017structural}, producing devices suitable for gas transport measurements.  Nanoscale imaging has shown the expected structure, with long-range crystalline order over areas of at least $300 \times \SI{300}{\nano\meter^2}$ and low defect density \cite{muller2018two,grossmann2021onsurface}.

Here we report both experimental measurements and atomistic molecular dynamics simulations of gas flow through free-standing poly(carboxyfantrip) membranes.  The permeances of five gases (\ce{H2}, \ce{He}, \ce{N2}, \ce{CO2}, and \ce{SF6}), spanning nearly a 100-fold range in molecular weight ($M_W$), are found to follow  Graham's law of effusion, scaling as $1/\sqrt{M_W}$. Surprisingly, we find both with experiment and simulation that the transport is consistent with molecular effusion with an effective open area independent of the size of the molecule.  This suggests the importance of interactions between the gas and the 2D material which increase the effective area compared to a non-interacting ballistic model.

\section*{Ideal Gas Permeance}
The ultimate upper bound on permeance is given by the rate $\Phi$ at which molecules of an ideal gas of molecular weight $M_W$ strike a surface.  From elementary kinetic theory, and assuming the molecules are point-like, one has 
\begin{equation}\label{eq:wallflux}
\Phi = \frac{nA\bar{v}}{4} = \frac{ P~A~N_A}{\sqrt{2\pi M_W RT}},
\end{equation}
where $n$ is the number density, $A$ is the area, $\bar{v} = \sqrt{8RT/\pi M_W}$ is the mean molecular speed, $ P$ is the pressure, $R$ is the gas constant, $N_A$ is Avogadro's number, and $T$ is the temperature.  For a porous membrane with  open area $A_p$, and gas on either side with pressure $P_L-P_R=\Delta P$, the net flux $\dot{N}$ of gas molecules that passes through the membrane is:
\begin{equation}\label{eq:effusion}
\dot{N} =  \frac{A_p}{A}\frac{1}{\sqrt{2\pi M_W RT}}\cdot \left(N_A A~\Delta P\right)=Q~\left(N_A A~\Delta  P\right).
\end{equation}
This defines the effusion permeance, $Q$, which represents an upper bound on transport through an atomically thin membrane in the free-molecular-flow regime.  

This regime applies when the Knudsen number $\text{Kn} = \lambda / d \gg 1$, where $\lambda$ is the mean free path and $d$ is the pore diameter. Experiments were performed at room temperature with gas pressure near \SI{1}{\bar}, and a small pressure difference, $\Delta P \simeq \SI{374}{\pascal}$, was applied across the membrane to drive flow.  At these conditions for nitrogen gas, $\lambda\approx \SI{63}{\nano\meter}$, easily satisfying the molecular flow requirements for a single pore with effective diameter of $\sim \SI{1.2}{\nano\meter}$ as shown in Fig.~\ref{fig:combined}(a). Furthermore, the simple effusion rate, $\dot{N}$, does not consider the effect of the size of the molecule.  Naively, one would expect the effective area of a pore to be reduced by the finite size of the molecule.

\section*{Experiment}\label{subsec:gastransport}

\begin{figure*}[!t] 
    \centering
\includegraphics[width=1.0\textwidth]{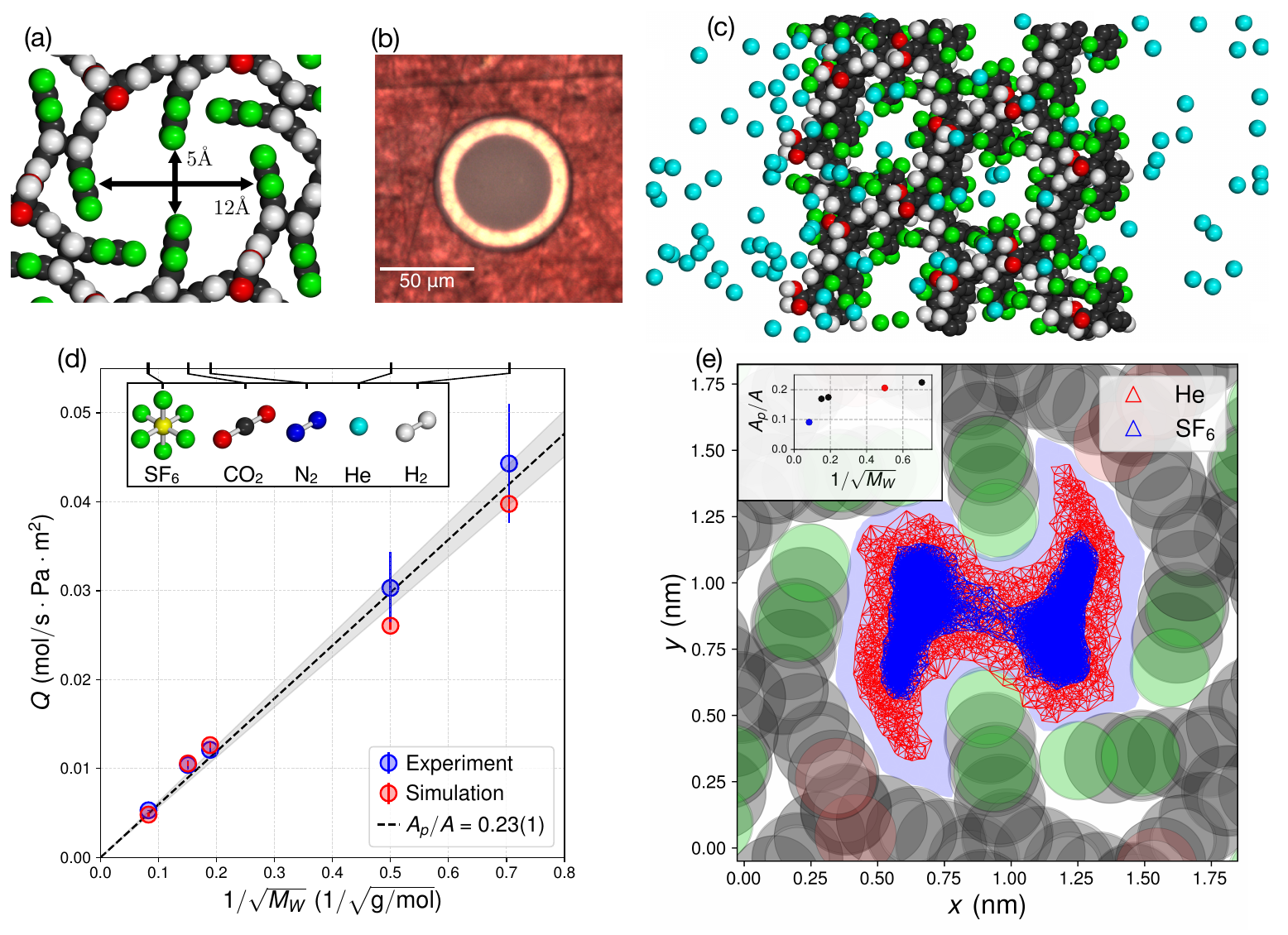}
    \caption{
        (a) A zoom-in of a pore of poly(carboxyfantrip). (b) A single-layer sheet of poly(carboxyfantrip) spanning a $\sim 50\mu\text{m}$ hole, imaged via polarized optical microscopy. (c) Snapshot of a molecular dynamics simulation of He gas flow through poly(carboxyfantrip). (d) The measured permeance of five gases as compared to molecular dynamics simulations. The dashed line is a Graham's law fit to the full set of experimental and simulation data, yielding an open area fraction $A_p/A=0.23(1)$. Reported uncertainties on simulation data are standard statistical errors, while experimental error bars correspond to the maximum of the standard deviation over multiple trials or, when there were only two trials, the difference between them.
    (e) Molecular dynamics determination of the open area $A_p/A$ from gas trajectories through a hole in poly(carboxyfantrip). The area is computed via a triangulation of crossing points for both He (red triangles) and SF$_6$, giving $A_p/A=0.21$ and $A_p/A=0.09$, respectively. A triangulation based on kinetic radii of the poly(carboxyfantrip) atoms (light blue shaded region) gives a significantly higher open area. The inset shows open area fractions computed via particle crossing triangulation for each of the molecular species.}
    \label{fig:combined}
\end{figure*}

To measure the rate of gas transport, free-standing poly(carboxyfantrip) membranes were suspended over single \SI{50}{\micro\meter} apertures in copper TEM grids (Fig.~\ref{fig:combined}b) and measured using a capillary bubble-flow apparatus (see Methods). The single-orifice geometry provided a well-defined area and made membrane failure easy to identify, since damaged films produced visibly defective samples or anomalously large gas fluxes. After optical verification of membrane integrity, permeance was measured at room temperature under a small applied pressure difference of  $\SI{374}{\pascal}$.   Gas flow was determined from the displacement of a soap-film slug in a calibrated micropipette, with repeated measurements used to confirm reproducible steady-state conditions. Measurements were performed for \ce{H2}, \ce{He}, \ce{N2}, \ce{CO2}, and \ce{SF6}, with traversal times varying from \SIrange{40}{350}{\second}, providing a direct experimental probe of molecular-size-dependent transport through the membrane with results shown in Fig.~\ref{fig:combined}d.  

Observations agree with Eq.~(\ref{eq:effusion}) and support that gas transport through poly(carboxyfantrip) proceeds near the ideal, wall-collision-limited rate with an open area fraction near $23\%$.  The observation that \ce{SF6} (kinetic diameter $\sim$\SI{0.55}{\nano\meter}) permeates through the membrane was unexpected given the estimated pore size.  This suggests that the pores of poly(carboxyfantrip) interact with the gas and may possess some conformational flexibility that allows molecules larger than the static pore diameter to transit.

\section*{Atomistic Simulations}\label{subsec:simulation}
To verify the experimental gas flow measurements and microscopically determine the open area fraction of the membrane, we employed the Large-scale Atomic/Molecular Massively Parallel Simulator (LAMMPS) \cite{LAMMPS} to perform molecular dynamics simulations of the gas permeance of poly(carboxyfantrip) with details included in the Methods section.  The permeance $Q$ was computed via an equilibrium setup, where equal numbers of gas molecules are initially placed on either side of the 2D polymer held at $z=0$ oriented in the $xy$-plane. During the simulation, gas molecules can transit the membrane and we record the total number of crossings as a function of time, which grows linearly with a slope proportional to the permeance.  We note that previous studies of nanoporous graphene \cite{Sun:2014ec} have verified that nonequilibrium simulations, which mimic gas flow experiments, give consistent results with the equilibrium setting considered here. 

\begin{figure*}[t]
\centering
\includegraphics[width=1.0\textwidth]{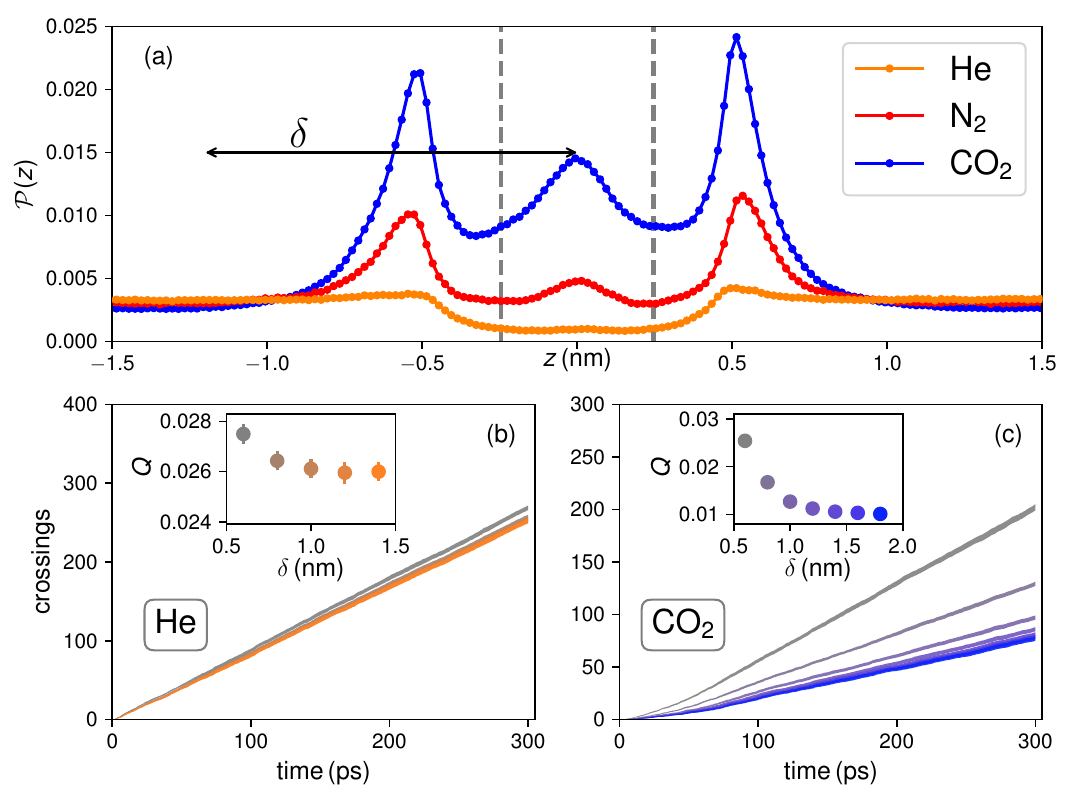}
\caption{$(a)$ The probability for gas molecules to be found a distance $z$ away from the plane of the poly(carboxyfantrip) sheet, computed from MD simulations of He, $\text{N}_2$, and $\text{CO}_2$. Dashed lines indicate the maximum and minimum $z$ coordinates of atoms in the poly(carboxyfantrip) sheet. $\text{N}_2$ and $\text{CO}_2$ demonstrate a significant attraction to poly(carboxyfantrip) at room temperature, requiring us to define gas crossings in terms of the parameter $\delta$. $(b,c)$ Total number of gas crossings as a function of time for different values of $\delta$. The slopes are used to compute the permeance $Q$ (mol /s $\cdot$ Pa $\cdot$ $\text{m}^{2}$), shown in the inset, which converges for sufficiently large $\delta$.} \label{fig:crossings}
\end{figure*}

The most important consideration in our permeance calculations is the length scale associated with the gas-poly(carboxyfantrip) interaction and resulting traversal process. In Fig. \ref{fig:crossings}a we show the probability of finding a gas molecule a distance $z$ away from the poly(carboxyfantrip) sheet. 
In each case there is a background suppression of density near $z=0$ due to the excluded volume of the poly(carboxyfantrip), but additionally we find that the molecular gases show a pronounced attraction to the surface and pore interior, leading to dynamic adsorption phenomena. As a result, we define a membrane crossing event in terms of a length $\delta$, where crossings are only recorded when molecules traverse the region from $z=-\delta$ to $z=\delta$ (or back).  We expect that $\delta$ should be larger than the range of the interaction potential between the gas molecules and the poly(carboxyfantrip) surface.

In Fig. \ref{fig:crossings}b, c we show the total number of gas molecule crossings as a function of time as $\delta$ is systematically increased.  Results are based on 20 independent simulations of He and $\text{CO}_2$, where ensemble results are used to compute a mean value and standard error on the number of crossings, depicted as thick shaded lines. In each case we perform a linear fit to the data, where we exclude early times before the system has reached equilibrium.  The slope of these lines is used to compute the permeance, defined as the slope divided by the pressure and the membrane area. The pressure is computed via the ideal gas law using the average number of molecules in the region $\lvert{z}\rvert > \delta$, valid in the low-density room temperature regime considered here. The calculated permeances ($Q$) as a function of $\delta$ are plotted in the insets with propagated error bars. For He we find a very mild dependence on the threshold parameter $\delta$, whereas for molecules with larger molecular weight,  we find strong dependence on $\delta$, with a clear convergence of the permeance when $\delta$ probes the bulk region far from the membrane.  


$\delta$-converged results for the permeance of each species are shown in Fig.~\ref{fig:combined}d where we find remarkable quantitative agreement with experiment.  Access to microscopic information on the location in the plane of crossings allows us to independently estimate the area fractions for each molecular species via triangulation as shown in Fig.~\ref{fig:combined}e with further information in Extended Data Fig.~\ref{fig:Hetraj}.  As seen in the inset, we find that the open area decreases as a function of increasing molecular weight, while the permeance remains proportional to $1/\sqrt{M_W}$.  Given that the gas molecules in our simulations have an appreciable interaction with the poly(carboxyfantrip) sheet, the agreement between the experiments and simulation with the non-interacting theory suggests an interaction-mediated transit mechanism.

\section*{Discussion}

The permeance values measured here can be compared with results for other atomically thin membranes, and in particular, nanoporous graphene membranes, that have been extensively studied for gas separation applications \cite{Sun:2014ec,Gharibzahedi2016,Wang2018}. It is conventional in the membrane separation literature to report transport values in terms of gas permeation units, where \(1~\mathrm{GPU}=3.35\times10^{-10}~\mathrm{mol\,m^{-2}\,s^{-1}\,Pa^{-1}}\). In these units, crack-free porous single-layer graphene membranes have reported \ce{H2} permeances of over $10^3$~{GPU} at high temperature \cite{Huang2018CrackFreeGraphene}, and permeances above $5\times 10^3$~{GPU} have been reported in both angstrom-scale etched graphene \cite{Zhao2019AngstromGraphene} and chemical vapor deposition of porous graphene \cite{Yuan2021PorousGrapheneCVD}.   More recently, nanofabricated \ce{CO2}-selective graphene membranes have achieved \ce{CO2} permeances near $10^4$~{GPU} \cite{Hsu2024PyridinicGrapheneCO2,Hao2025ScalableCO2Graphene}.  Simulations of a porous 2D polymer membrane have predicted \ce{CO2} permeances of \num{3e5}~GPU \cite{Schrier:2012ca}.  

Converting to GPUs, poly(carboxyfantrip) yields comparatively giant values,
ranging from $Q_{\ce{H2}} \simeq 1.3\times10^{8}~\mathrm{GPU}$ to 
$Q_{\ce{SF6}} \simeq 1.6\times10^{7}~\mathrm{GPU}$, 3--4 orders of magnitude
larger than most atomically thin gas-sieving membranes. The observed transport operates near the ideal effusive limit for an atomically thin porous membrane where the permeance is set by the molecular collision rate with the pore opening and by the open area fraction $A_p/A$. Permeances approaching this magnitude have previously been reached only with  graphene membranes perforated by nearly a million, much larger ($\sim \SI{10}{\nano\meter}$), serially fabricated ion-milled pores \cite{Celebi:2014sc}. This is in contrast to the work here where the pores are of molecular size, structurally identical, with area fraction approaching unity, and created during the material synthesis.

For separation applications, mixed-gas measurements provide a more realistic assessment of membrane performance under practical operating conditions. Studies on nanoporous graphene have shown that mixed-gas selectivities can differ significantly from single-gas selectivity ratios, for example due to competitive adsorption \cite{Yuan2018,Huang2018CrackFreeGraphene}. Future work should investigate mixed-gas permeation through poly(carboxyfantrip) to assess its practical separation performance.

The unexpectedly high \ce{SF6} permeance is an important constraint on the transport mechanism. If the pores acted as rigid, hard apertures smaller than the kinetic diameter of \ce{SF6}, strong suppression of the permeance of \ce{SF6} would be expected.  Instead, \ce{SF6} follows the same approximate $1/\sqrt{M_W}$ trend as the smaller gases, suggesting the possibility of interaction effects, dynamic fluctuations of the pore rim, rare larger transport pathways, or some combination of these. Recent studies of angstrom-scale graphene pores illustrate the distinction. True molecular sieving can produce activated, exponentially selective transport when pores approach the molecular kinetic diameter \cite{Sun2021AngstromPores}.  Conversely, pore-edge dynamics and molecule-pore interactions can eliminate simple effusive behavior and produce selective gating, as shown for \ce{CO2} permeation through dynamic graphene pores \cite{Bondaz2025:gc02}. In poly(carboxyfantrip), our results point to the opposite conclusion: adsorption and flexibility may be present microscopically, but the net single-gas flux remains close to the effusive limit.

Further investigation through variable-temperature measurements and molecular
dynamics simulations with flexible pore models would help clarify the transport
mechanism for larger molecules. As poly(carboxyfantrip) selectivities are
essentially limited by Graham's law, the next challenge is to determine whether
chemical modification, multilayer stacking, or
controlled lattice distortion can add selectivity without destroying the large
open-area advantage.  Demonstration of isotopic separation of gases, such as $^3$He/$^4$He mixtures, through multiple barriers of poly(carboxyfantrip) would also be a possible next step.

Lastly, there are other areas of science which require the structure and transport properties demonstrated here.  Josephson junctions for superfluid $^4$He require an array of nm-scale pores \cite{cheishvili1965analog,sato2012superfluid}, and this opens the possibility for very high sensitivity quantum sensors and devices for use in both fundamental science and practical application \cite{ellers2026quantum}.

\section*{Summary}\label{sec:conclusion}

We have measured and simulated gas permeance through free-standing membranes of poly(carboxyfantrip), a crystalline, nanoporous two-dimensional polymer with atomically identical pores. The measured permeances for five gases (\ce{H2}, \ce{He}, \ce{N2}, \ce{CO2}, and \ce{SF6}) follow Graham's law of effusion, scaling as $1/\sqrt{M_W}$, and approach the ideal gas wall-collision limit predicted by kinetic theory. Molecular dynamics simulations quantitatively reproduce experimental permeances within reported uncertainties.  The agreement between experiment and simulation allows for an estimate of the open area of the membrane of $A_p/A = 0.23(1)$ within the non-interacting theory.   This value is consistent with the open area determined via particle trajectory triangulation for light, weakly interacting gases, demonstrating that transport through poly(carboxyfantrip) is governed by effusion through the intrinsic pores of the material.   These results establish bottom-up, synthetic 2D polymers as a viable platform for giant gas flow through atomically thin membranes, which 
motivates future studies of functionalization with the goal of mixed-gas selectivity.

\section*{Methods}\label{sec:methods}

\subsection*{Fabrication and Characterization}\label{subsec:fabrication}

Poly(carboxyfantrip) films were prepared at the air--water interface \cite{Murray:2016thesis,muller2017structural} by spreading a \SI{0.2}{\milli\gram\per\milli\liter} solution of the carboxyfantrip monomer in 10:1 chloroform:isopropanol on a Langmuir trough, compressing to \SI{10}{\milli\newton\per\meter}, and irradiating with a \SI{5}{\watt}, \SI{365}{\nano\meter} LED for \SI{30}{\minute}.  The surface-pressure/mean-molecular-area isotherm gives a mean molecular area of \SI{175}{\angstrom^2\per molecule}, consistent with the area calculated from the single-crystal structure of the non-amphiphilic analog, fantrip \cite{Kissel:2014nc}. 

Langmuir films were prepared on a NIMA 312 trough with two barriers; surface pressure was monitored with a paper Wilhelmy plate. \SI{20}{\micro\liter} of the monomer solution was spread on Milli-Q water and, after \SI{30}{\minute} for solvent evaporation, compressed at \SI{3}{\milli\meter\per\minute}. The LED (Led-Engin LZC-70U600) was positioned $\sim$\SI{13}{\centi\meter} above the interface.

Transfer to copper TEM grids (Gilder GA50, single \SI{50}{\micro\meter} hole) was accomplished by Langmuir--Schaefer deposition: grids were placed horizontally onto the irradiated film and removed using a home-built Delrin basket raised at \SI{1}{\milli\meter\per\minute}; grids were allowed to dry for \SI{2}{\hour} before removal.  Transfer to \ce{SiO2} for optical and AFM characterization was performed by immersing the substrate at \SI{45}{\degree} before spreading and pulling through the polymerized film at \SI{1}{\milli\meter\per\minute}.

Optical microscopy of films transferred to \ce{SiO2} shows large-area ($> \SI{1}{\milli\meter^2}$), homogeneous, defect-free membranes. The transferred polymer films exhibit straight cracks with corner angles of $\sim$\SI{70}{\degree} and $\sim$\SI{110}{\degree}, consistent with the unit cell angles of the non-amphiphilic poly(fantrip) crystal structure, indicating long-range crystalline order (Extended Data Fig.~\ref{fig:cracks}). Atomic force microscopy (AFM) of the monomer film on \ce{SiO2} gives a step height of $\sim$\SI{1.3}{\nano\meter}, confirming that the film is a monolayer.

High-resolution AFM (HR-AFM) phase imaging of poly(carboxyfantrip) on mica reveals a distorted hexagonal array of pores with unit cell parameters $a = \SI{1.8 \pm 0.3}{\nano\meter}$, $b = \SI{2.0 \pm 0.3}{\nano\meter}$, $\gamma = \SI{71 \pm 3}{\degree}$ (Extended Data Fig.~\ref{fig:hrafm}).  These values are in good agreement with the single-crystal X-ray diffraction data for poly(fantrip) ($a = \SI{1.71}{\nano\meter}$, $b = \SI{1.92}{\nano\meter}$, $\gamma = \SI{70}{\degree}$) \cite{Kissel:2014nc}.  The domain size could not be determined because the lattice cannot be resolved at scan sizes larger than $\sim$\SI{100}{\nano\meter}.

Scanning electron microscopy (SEM) confirms that monolayer poly(carboxyfantrip) films can span $25 \times \SI{25}{\micro\meter^2}$ holes on TEM grids, demonstrating the mechanical coherence of the membrane.  For gas transport measurements, free-standing membranes were prepared by Langmuir--Schaefer transfer onto single-hole copper TEM grids (Gilder GA50, \SI{50}{\micro\meter} hole diameter). Film quality was verified by optical microscopy prior to each measurement to confirm that the membrane fully spanned the hole without visible tears or folds.

\subsection*{Characterization}\label{subsec:methods_characterization}

Optical microscopy was performed using a Leica DM2500P polarizing microscope.  AFM imaging was performed in tapping mode on an Asylum Research Cypher ES using Budget Sensors Tap150Al-G probes (resonance frequency \SI{150}{\kilo\hertz}, spring constant \SI{5}{\newton\per\meter}).  High-resolution AFM was performed on the same instrument using Olympus AC55TS probes (resonance frequency \SI{1600}{\kilo\hertz}, spring constant \SI{85}{\newton\per\meter}) in tapping mode under dry \ce{N2}; imaging was performed in the repulsive regime, as the polymer lattice could not be resolved in the attractive regime.  

\subsection*{Gas Permeance Measurements}\label{subsec:methods_gasflow}

Gas flux measurements were performed using a custom-built capillary bubble flow meter apparatus as shown in Extended Data Fig.~\ref{fig:apparatus}. The TEM grid bearing the poly(carboxyfantrip) membrane was sealed into a sample holder connected to a gas reservoir. The system was purged with the test gas for at least \SI{30}{\minute} prior to measurement.  A differential pressure of \SI{1.5}{in\,\ce{H2O}} ($\simeq \SI{374}{\pascal}$) was applied across the membrane, monitored by a differential pressure gauge (Heise, 0--50 in \ce{H2O}). Gas permeating through the membrane displaced a soap-film slug in a calibrated \SI{25}{\micro\liter} glass micropipette (Drummond Wiretrol), and the transit time was recorded manually. On average ten trials were performed for each gas, with at least two successful transits to ensure steady-state conditions.  Permeance $Q$ was calculated from the measured volumetric flow rate, and the membrane area according to Eq~(\ref{eq:effusion}). 
Measurements were performed at room temperature ($\sim$\SI{295}{\kelvin}) for \ce{H2}, \ce{He}, \ce{N2}, \ce{CO2}, and \ce{SF6}.

For gas transport measurements, free-standing films of poly(carboxyfantrip) were transferred onto copper TEM grids (Gilder GA50, having a single \SI{50}{\micro\meter} hole diameter).  With great care, free-standing films could be obtained in low yield, about one in twenty attempts. The quality of each film was assessed by optical microscopy prior to measurement to confirm that the film spanned the hole without visible tears or folds.  The diameter of the hole was measured by optical microscopy.  The use of a single orifice provided a binary outcome in the experiment, and damaged membranes were evident from optical inspection or anomalously high gas flows (see Extended Data Fig.~\ref{fig:ruptured}).

\subsection*{Molecular Dynamics Simulations}\label{subsec:methods_md}

All-atom molecular dynamics simulations were performed using LAMMPS \cite{LAMMPS} where the universal force field (UFF) \cite{Rappe1992} was taken as a minimal model of the interatomic potentials. In our simulations, poly(carboxyfantrip) was placed at the center of the box ($z=0$), with periodic boundaries in the $(x,y)$ direction. The number of unit cells in the $x$ and $y$ direction is chosen so that the poly(carboxyfantrip) sheet wraps around the boundaries with minimal distortion. In order to hold the sheet in place during the simulation, the bridgehead carbon atoms are fixed in place, while the rest of the poly(carboxyfantrip) sheet evolves according to the NVT ensemble (fixed particle number, volume and temperature) of LAMMPS.  The gas is confined to either side of the poly(carboxyfantrip) sheet by reflective walls at $z_{\text{min}}=-150$ \AA ~and $z_{\text{max}}=150$ \AA. The poly(carboxyfantrip) sample spans a length of $\Delta x = 51.3158$ \AA~ and width of $\Delta y = 35.324$ \AA ~and contains six pores. We have checked that our permeance calculations do not depend on the volume, number of gas molecules, pressure, or poly(carboxyfantrip) sample size.

For each gas species (\ce{H2}, \ce{He}, \ce{N2}, \ce{CO2}, \ce{SF6}), 20 independent simulations were performed with equal numbers of gas molecules initially distributed on both sides of the sheet. The total number of crossings through the membrane was recorded as a function of time, and the permeance was computed from a linear fit to the crossing data, as described in the main text.

\begin{acknowledgments}
We thank the staff at the Molecular Foundry for assistance with atomic force microscopy. Work at the Molecular Foundry was supported by the Office of Science, Office of Basic Energy Sciences, of the U.S. Department of Energy under Contract No. DE-AC02-05CH11231.
\end{acknowledgments}

\section*{Declarations}

\begin{description}
\item[Funding] KCS, BTK, AD, and JD acknowledge funding from NSF QuSeC-TAQS program through grant number 2326801.  
\item[Competing interests] The authors declare no competing interests.
\item[Data availability] The experimental permeance data supporting the findings of this study are available within the paper and its Supplementary Information. The raw molecular dynamics data have been deposited on Zenodo at Ref.~\cite{datarepo}. 
\item[Code availability] The LAMMPS input files and the analysis scripts used to generate the figures in this paper are available at Ref.~\cite{repo}.
\item[Author contributions] JD and AD designed and performed numerical simulations and data analysis. DJM and WBT performed synthesis, device fabrication, and transport measurements.  JCH contributed to preliminary simulations. KCS contributed to data analysis and project organization. BTK conceived the project and supervised experimental work and preliminary simulations. All authors contributed to writing the manuscript.
\end{description}

\section*{Additional information} Supplementary Information is available for this paper. Correspondence and requests for materials should be addressed to Adrian Del Maestro or Benjamin T. King.

\bibliography{references_arxiv}

\clearpage
\onecolumngrid
\renewcommand\figurename{Extended Data Fig.}
\setcounter{figure}{0}
\renewcommand{\theHfigure}{ED.\arabic{figure}}

\begin{figure}[!htbp] 
    \centering
\includegraphics[width=0.6\textwidth]{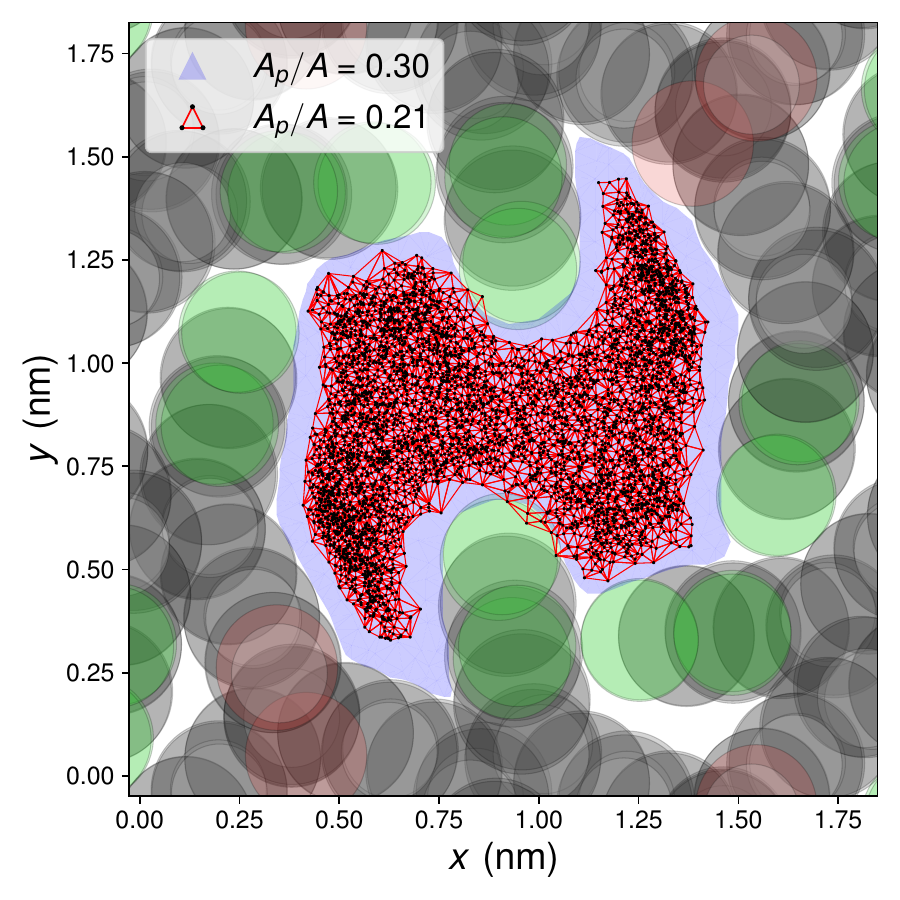}
\caption{The open area computed via an alpha shape triangulation (red triangles) of gas molecule crossings (black dots), shown here for a simulation of helium. The resulting triangulation gives an open area of $A_p/A=0.21$.  The open area based on kinetic radii of poly(carboxyfantrip) atoms (light blue) gives a significantly higher open area estimate of $A_p/A=0.30$.}
    \label{fig:Hetraj}
\end{figure}
\begin{figure}[!htbp]
    \centering
    \includegraphics[width=\textwidth]{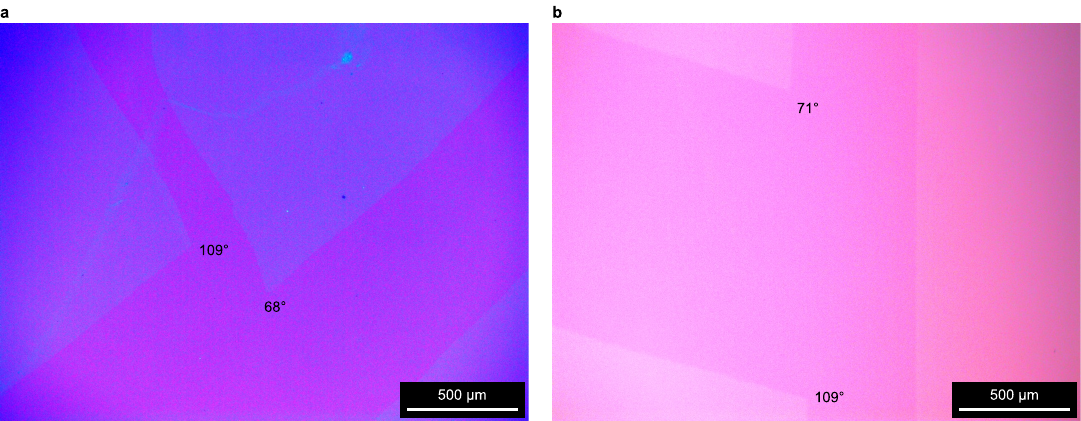}
\caption{\textbf{Crystallographic fracture of poly(carboxyfantrip) films.} \textbf{a}, \textbf{b}, Optical micrographs of two monolayer films transferred to \ce{SiO2}/Si. The films break along straight cracks that meet at corners of $\sim$\SI{70}{\degree} and $\sim$\SI{110}{\degree} (labelled), matching the \SI{70}{\degree} unit-cell angle of poly(fantrip) \cite{Kissel:2014nc} and indicating long-range crystalline order over areas of \si{\milli\meter^2}. Contrast was adjusted by a single linear stretch applied to the whole of each image. Scale bars, \SI{500}{\micro\meter}.}
    \label{fig:cracks}
\end{figure}
\begin{figure}[!htbp]
    \centering
    \includegraphics[width=\textwidth]{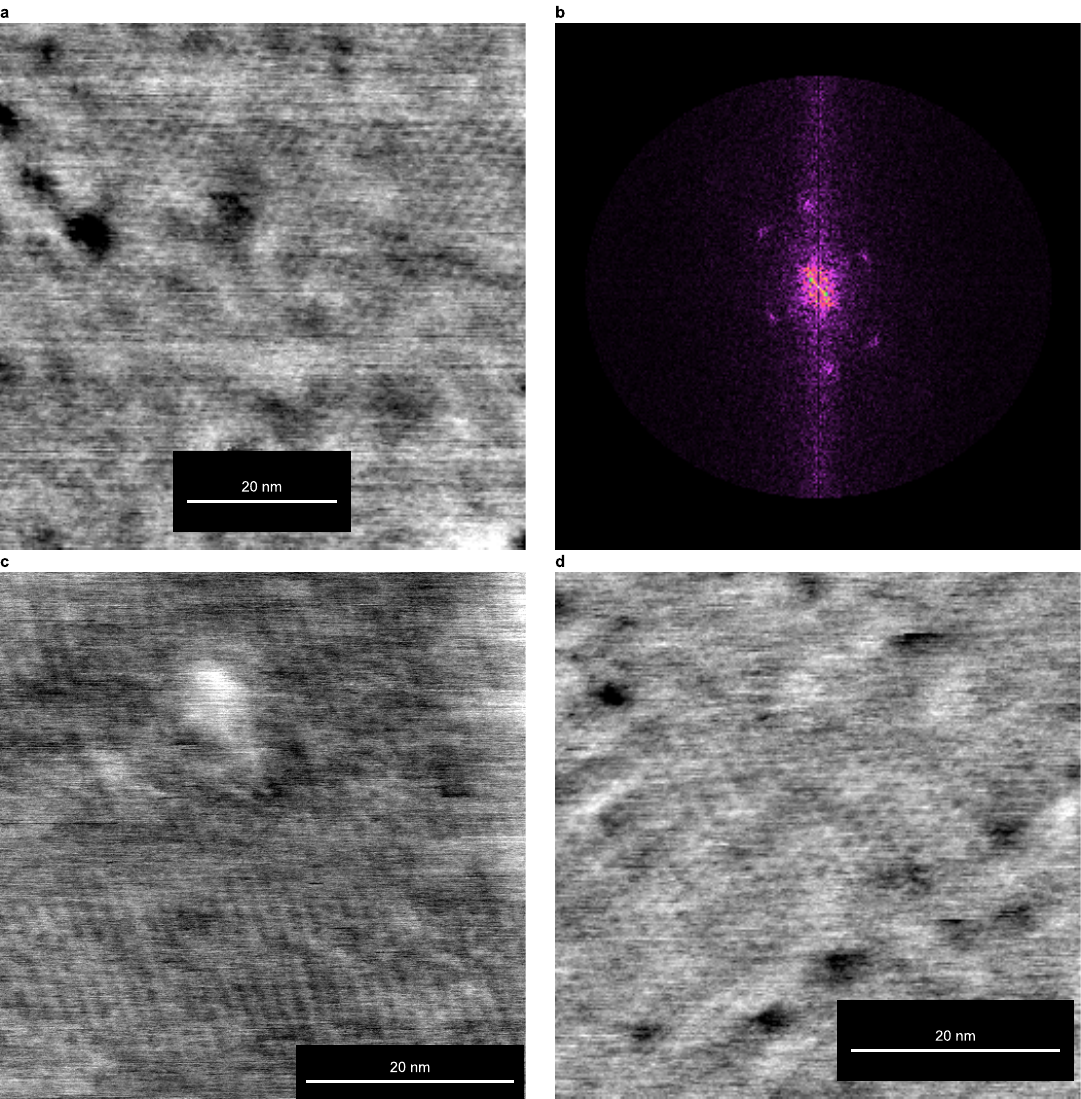}
\caption{\textbf{High-resolution AFM of poly(carboxyfantrip) on mica.} \textbf{a}, Phase image ($\SI{70}{\nano\meter} \times \SI{70}{\nano\meter}$) of a monolayer transferred to freshly cleaved mica, showing a distorted hexagonal array of pores. \textbf{b}, Two-dimensional fast Fourier transform of \textbf{a}. \textbf{c}, \textbf{d}, Phase images ($\sim \SI{50}{\nano\meter} \times \SI{50}{\nano\meter}$) of other regions. Images were acquired at the Molecular Foundry on an Asylum Research Cypher ES in tapping mode in the repulsive regime under dry nitrogen with Olympus AC55TS probes; the lattice could not be resolved in air or in the attractive regime. Analysis of multiple images from several samples gives $a = \SI{1.8 \pm 0.3}{\nano\meter}$, $b = \SI{2.0 \pm 0.3}{\nano\meter}$ and $\gamma = \SI{71 \pm 3}{\degree}$. Dark spots are regions of damped phase over mounds a few \si{\angstrom} high, attributed to contaminants trapped beneath the film. Scale bars, \SI{20}{\nano\meter}.}
    \label{fig:hrafm}
\end{figure}

\begin{figure}[!htbp] 
    \centering
\includegraphics[width=1.0\textwidth]{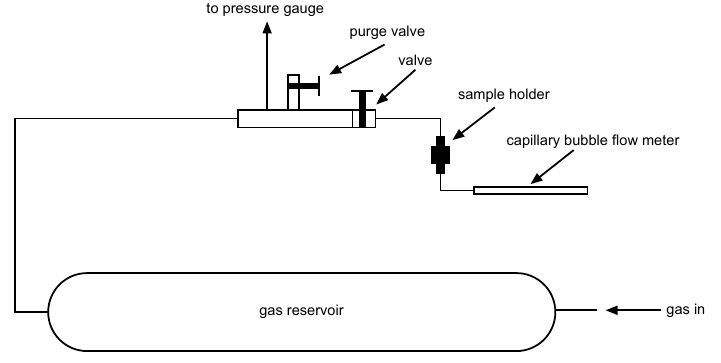}
\caption{Schematic of gas flow apparatus}
    \label{fig:apparatus}
\end{figure}

\begin{figure}[!htbp] 
    \centering
    \includegraphics[width=0.4\textwidth]{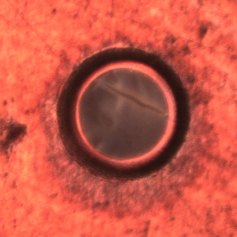}
\caption{Optical evidence of a ruptured film imaged via polarized optical microscopy.}
    \label{fig:ruptured}
\end{figure}

\clearpage
\onecolumngrid
\ifarXiv
    \pdfximage{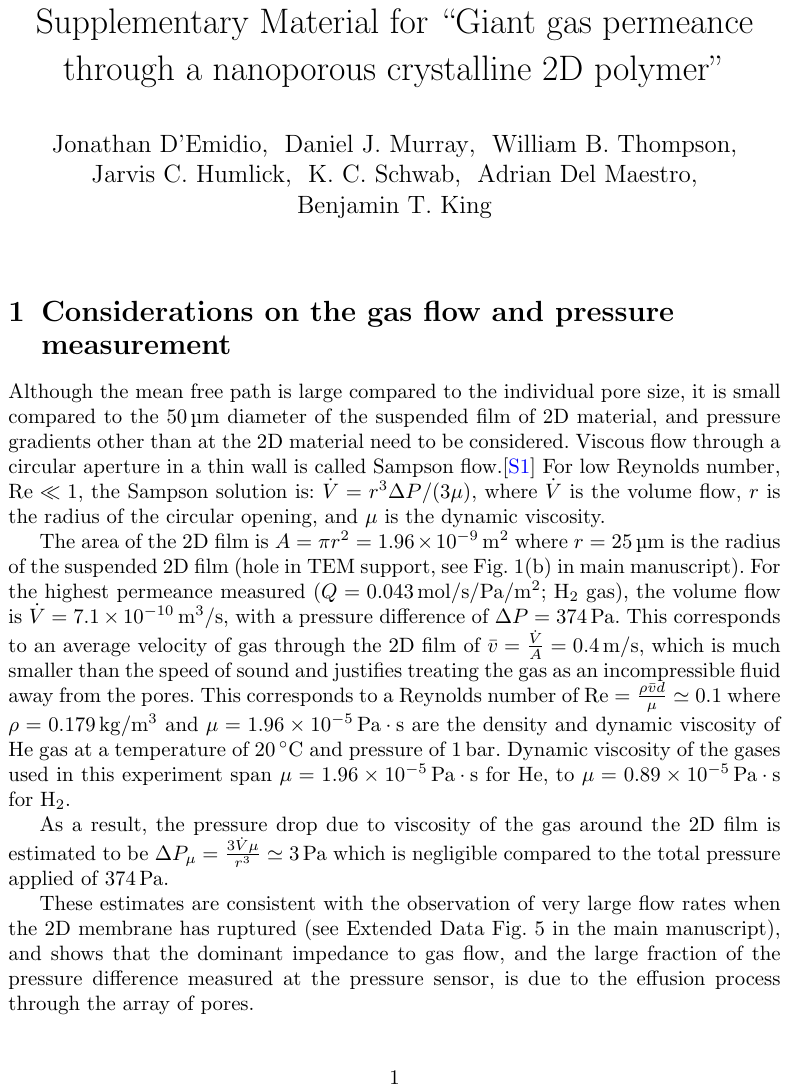}
    \edef\numbersupplementpages{\the\pdflastximagepages}
    \foreach \supplementpage in {1,...,\numbersupplementpages}{%
        \clearpage
        \includepdf[pages=\supplementpage,pagecommand={\thispagestyle{empty}},offset=0 0]{\supplementfilename.pdf}
    }
    \clearpage
\fi

\end{document}